\documentclass[conference]{IEEEtran}

\usepackage[hyphens]{url}
\usepackage{float}
\usepackage{cite}
\usepackage{amsmath,amssymb,amsfonts}
\usepackage{algorithmic}
\usepackage{graphicx}
\usepackage{textcomp}
\usepackage{xcolor}
\def\BibTeX{{\rm B\kern-.05em{\sc i\kern-.025em b}\kern-.08em
    T\kern-.1667em\lower.7ex\hbox{E}\kern-.125emX}}

\begin{document}
\bstctlcite{IEEEexample:BSTcontrol}

\title{Security Risk Analysis Methodologies for Automotive Systems}

\author{\IEEEauthorblockN{Mohamed Abouelnaga, Christine Jakobs}
}

\maketitle

\thispagestyle{plain}
\pagestyle{plain}

\begin{abstract}
Nowadays, systematic security risk analysis plays a vital role in the automotive domain. The demand for advanced driver assistance systems and connectivity of vehicles to the internet makes cyber-security a crucial requirement for vehicle manufacturers. This paper summarizes the risk analysis method stated in the recently released automotive security standard ISO/SAE 21434 \cite{1}, which lays the high-level principles for threat analysis and risk assessment (TARA) methods. Following, we introduce a specific use case to compare different security analysis approaches which OEMs can benefit from to achieve compliance with the standard.
\end{abstract}


\section{Introduction}
The constant increase of software functionalities for vehicles to satisfy consumers' demand and the current trend of developing connected vehicles and autonomous driving led to the development of more advanced systems. This requires a connection with the outside world of the vehicle, either the infrastructure or the other road vehicles. Also, the connection to the internet became an essential part of the infotainment systems. Therefore, the vehicles are not closed systems anymore. The urgency to deal with the cybersecurity attacks and vulnerabilities \cite{2}, \cite{3}, \cite{4}, \cite{5} has become as important as developing new functionalities \cite{6} as the violation of security goals is no more affecting data only. It may lead to financial problems, affect the reputation of the manufacturers, or even affect the safety of humans \cite{7}, \cite{13}.

An essential step toward building security-aware systems is obtaining security requirements from threat modeling and risk assessment. In risk analysis, we analyze different attack possibilities and calculate their impact on various categories. This process also enables better testing and validation as we can use attack paths as test cases.

However, the complexity of functionalities has needed a more systematic way to conduct the risk analysis. Although different organizations have developed different standards to serve the automotive field, such as ISO 26262 \cite{8} for safety, there was still a need for a common standard for security. That has led the International Organization for Standardization (ISO) and the Society of Automotive Engineers (SAE) to work on ISO/SAE 21434 \cite{1} since 2016 and successfully release it in August 2021. 

This standard deals with high-level steps to be applied in risk analysis; however, it does not specify use cases or how to automate this process. As a result, many projects were already working on the problem of risk analysis for different use cases, such as EVITA \cite{9}, \cite{10}, and SecForCARS \cite{11}.

This paper will summarize the TARA method as stated in the standard. Also, we introduce a use case to compare both EVITA \cite{9} and HEAVENS \cite{10} approaches. 

Moreover, we will see how the SecForCARS \cite{11} project participated in building a well-organized database for attack paths.

\section{THREAT ANALYSIS AND RISK ASSESSMENT (TARA)}
We present a detailed overview of high-level activities that we should conduct in the context of risk analysis.
\subsection{Prerequisite}
The standard introduces the steps we should use to systematically tackle the risk analysis problem. However, it only presents WHAT, not HOW. For example, it does not provide what different values exactly mean and how we should map them for impact and attack feasibility ratings.

We carry out TARA steps after defining the Target of Evaluation (TOE) or item definition. This step includes:
\begin{itemize}
\item Item boundary: it distinguishes the item from other internal or external items to the vehicle and defines the interfaces between the item and the other items
\item Item functions: this describes the item's behavior during different phases (concept, development, production, maintenance)
\item Preliminary architecture: this describes the various components of the item, their connections, and external interfaces of the item
\item Assumptions: relevant information regarding the security assumptions, e.g., using encrypted messages
\end{itemize}
\subsection{TARA activities}
According to ISO/SAE 21434 \cite{1}, TARA comprises six activities, as shown in Fig. 1; each has inputs, different artifacts as its outputs, and recommended methods:

\textbf{\textit{Asset identification}}: an asset is any resource that has value. In a vehicle, assets can be in-vehicle devices such as ECUs, sensors and actuators, applications running on in-vehicle devices, and communication data.
\begin{figure}[htbp]
\centering
\includegraphics[width=\linewidth]{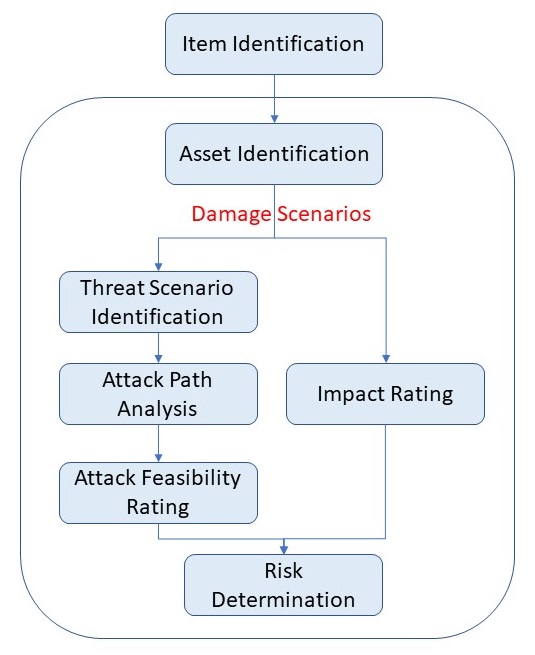}
\caption{TARA Activities}
\end{figure}

We can identify assets using the preliminary architecture and the assumptions obtained from the item identification activity. 

We can infer the damage scenarios from asset identification by associating the asset with specific cybersecurity properties. The ISO/SAE 21434 \cite{1} deals with the following properties: 
\begin{itemize}
\item Confidentiality: data must not be revealed to unauthorized parties
\item Integrity: data is complete and intact, so it should not be modified unauthorizedly or accidentally
\item Availability: data or system must be accessible when needed
\end{itemize}

However, in practice, we can use the following properties, e.g., when dealing with STRIDE \cite{12}:
\begin{itemize}
\item Non-repudiation: protection against an entity that falsely denies performing a particular action
\item Authenticity: proof of identity to verify that information originates from the claimed source
\item  Authorization: protection against privilege elevation through unauthorized access
\end{itemize}

For example, an asset would be data communication for brakes, and its cybersecurity property is integrity. The damage scenario is a collision with another vehicle when traveling at high speed resulting from unintended full brakes.

\textbf{\textit{Threat scenario identification}}: the threat scenario is the potential cause of compromise of assets' cybersecurity properties, which leads to the damage scenarios.

Threat scenarios can be identified either by group discussion of experts of different domains or by using systematic threat modeling approaches such as EVITA \cite{9}, TVRA \cite{14}, PASTA \cite{15}, and STRIDE \cite{12}.

For example, spoofing of CAN messages for brakes ECU leads to loss of integrity of those messages and thereby the loss of integrity of the braking functionality. 

\textbf{\textit{Impact rating}}: we assess damage scenarios against potential consequences for road users in four different categories: safety (S), financial (F), operational (O), and privacy (P). We can add more categories with a suitable justification.

The impact rating for each category has to be one of four values: "severe," major," moderate," or "negligible." We have to derive the safety rating from ISO 26262-3:2018 \cite{8}.

\textbf{\textit{Attack path analysis}}: besides the preliminary architecture and predefined environmental assumptions, we analyze the threat scenarios to identify the attack paths. We can achieve attack path identification by either top-down approaches such as attack trees, which analyze each threat scenario to deduce attack paths that realize it, or bottom-up approaches using vulnerability or weakness analysis.

\textbf{\textit{Attack feasibility rating}}: we should assess each attack path according to four categories:
\begin{itemize}
\item	High: if the attack path utilizes a low effort
\item	Medium: if the attack path utilizes a medium effort
\item	Low: if the attack path utilizes a high effort
\item	Very low: if the attack path utilizes a very high effort
\end{itemize}

This rating should be determined using one of the following approaches:
\begin{itemize}
\item	Attack potential-based approach: as defined in ISO/IEC 18045 \cite{23}, it measures the effort needed for successfully performing the attack. That relies on the potential of the attacker and used resources. Five core factors determine it: 
\begin{itemize}
\item	Elapsed time: the time required to identify the vulnerability and perform a successful attack
\item	Knowledge of the item or the component: acquired by the attacker
\item	Attacker expertise: related to the skill and the experience of the attacker 
\item	The window of the opportunity: related to the access conditions as access type, whether it is physical or logical, and the access time for the attacker to perform a successful attack 
\item	The equipment: available to the attacker to discover the vulnerability and perform the attack
\end{itemize}
\item	Attack vector-based approach: we can determine the feasibility rating according to the logical and physical distance between the attacker and the item or the component; the more remote, the higher rating \cite{22}
\item	Common Vulnerability Scoring System (CVSS) \cite{22}: we can define the feasibility rating by the exploitability metric (E), which depends on attack vector (V) (ranging from 0.2 to 0.75), attack complexity (C) (ranging from 0.44 to 0.77), the privileges required (P) (ranging from 0.27 to 0.85), and the user interaction (U) (ranging from 0.62 to 0.85). The proposed formula for calculating the exploitability value is as follows: 
\begin{equation}
E=8.22 \times C \times V \times P \times U \label{eq1}
\end{equation}
\end{itemize}

\textbf{\textit{Risk value determination}}: The risk of a threat scenario can be determined using two parameters, the likelihood of its attack path (consider the maximum feasibility rating in case of many associated attack paths). And the impact of the associated damage scenario (if more than one damage scenario results from the threat, we consider a different risk value for each damage scenario). The risk value is an integer ranging from "1" (minimum risk) to "5" (maximum risk). This value can be determined using user-defined formulas or risk matrices (the row rep-resents impact rating; the column represents the attack feasibility rating or vice versa).

\section{TARA IN PRACTICE}
Although several security models are concerned with threat analysis and risk assessment, they do not focus on the automotive field. For example, Threat, Vulnerability and Risk Analysis (TVRA) \cite{14} deals with security threats in the telecommunications domain, and Open Web Application Security Project (OWASP) \cite{25} focuses on web application security. 

Specific and detailed use cases would be a good approach for identifying threats and security requirements. They reflect on how to realize the users' goals by building scenarios describing the required system's functional properties. Use cases should focus on what the system should do instead of how to be done.

We present Road Speed Limit (RSL) use case to two different risk assessment frameworks, which are dedicated to identifying security issues for vehicles, EVITA \cite{9} and HEAVENS \cite{10}. The well-defined assessment in both frameworks and the impact on autonomous driving are the main reasons behind the choice of this use case. Also, applying the same use case to both approaches would clarify the differences.
 
Also, we use attack trees in both models for attack path analysis. The tree's root represents the attack's goal, and the leaves represent the ways to achieve this goal. Attack trees can contain AND- or OR- nodes to model the dependencies among the steps of the attack path \cite{21}.
\subsection{RSL use case} 
The road speed limit (RSL) restricts the vehicle's speed. Different parties can impose the speed limit, e.g., an authority or a fleet owner. According to EVITA \cite{9}, RSL is directly related to traffic information from/to other entities. An entity can be another car, a traffic light, or a roadside unit. This process includes exchanging traffic messages through Communication Unit (CU) between entities and collecting data through in-vehicle sensors, as shown in Fig. 2.

\begin{figure}[ht]
\centering
\includegraphics[width=\linewidth]{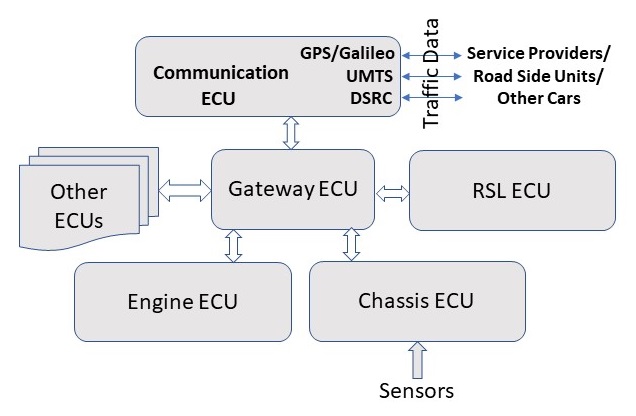}
\caption{Architecture of traffic information Exchange}
\end{figure}

\subsection{EVITA model} 
The E-safety vehicle intrusion protected applications (EVITA) \cite{9} project designed an efficient security solution for automotive onboard networks. Therefore, it introduces a risk analysis framework to protect security-relevant components against attacks.
\subsubsection{Asset identification} 
The possible affected assets are the wired infrastructure, roadside units, wireless communication data, and authorization keys.

\subsubsection{Threat scenarios and attack paths} 
Before proceeding with threat identification, we can consider EVITA \cite{9} as an attacker-centric approach, as it pays attention to attacker goals when deriving the possible threats.
Categories for possible attack motivations are as follows:
\begin{itemize}
\item Harming the driver or the passengers: does not have to be physical damage. The attacker may want to implicate the driver in legal troubles, e.g., by manipulating speed limits that do not necessarily lead to an accident but may result in fines that legally and financially harm the driver
\item	Gaining a reputation as a hacker: the attacker's primary goal is not to harm the system or the users but rather to publish the results of a successful attack to gain a reputation
\item	For financial gain: for example, the attacker may tamper with the vehicle for insurance fraud; he attacks the steering or brakes of another vehicle to provoke an accident
\item	To gain personal (non-financial) advantages: for example, going faster in the traffic, e.g., switching all traffic lights to green or directing other vehicles to alternative routes to make the way clear in front of the attacker
\item	Gain information about the manufacturer: one purpose can be that an attacker or another competing manufacturer wants to reveal the technical specifications of a specific item. Another goal can be destroying the reputation of a particular manufacturer, e.g., manipulating the safety to harm random users or violating the users' privacy
\item	Harm to the economy: attacking the infrastructure, which may lead to accidents, generate traffic jams, or disrupt the normal state of roads
\item	Mass terrorism: most of the attacks to achieve this are
similar to other categories but with a large scale of impact. As the attacker belongs to an organization, the availability of resources would not be a problem
\end{itemize}

Regarding modeling the threats, we use attack trees. As shown in Fig. 3, an attack tree has the following structure: the tree's root (level 0) represents an abstract attack goal. Its child nodes (level 1) describe the attack objectives satisfying the attack goal. We can estimate the severity of the attack's result at this level as these objectives harm the stakeholders (vehicle and other road users, authorities, service operators, and vehicle manufacturers). We can introduce some attack methods (level 2) to achieve each attack objective. Each attack method is composed of (AND/OR) logical combinations of attacks against assets known as "asset attacks, " representing the tree's leaves.

\begin{figure}[ht]
\centering
\includegraphics[width=\linewidth]{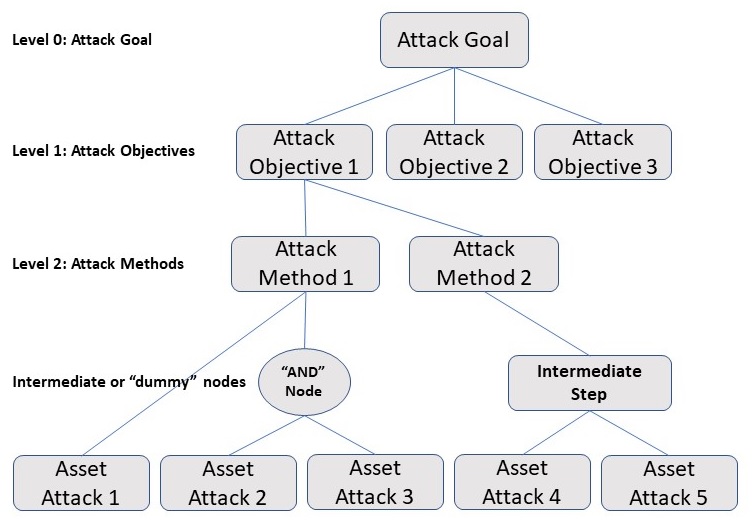}
\caption{Generic Attack Tree Structure \cite{9}}
\end{figure}

One attack goal for RSL is manipulating the speed limits. Two possible objectives can be: slowing down other vehicles behind the attacker by tampering with the infrastructure to be notified falsely of lower speed limits or increasing the speed limit by attacking the speed limit enforcing equipment (e.g., radars). The attack tree based on this goal is shown in Fig. 4.

\subsubsection{Impact rating}     
We assess the attack objectives against four categories: safety, financial, operational, and privacy. Nevertheless, we use five values ("0"," 1"," 2"," 3"," 4") to assess the severity of each category instead of four, as shown in the standard. 

We map the rating for the four categories as the following:

For the safety category $(S_S)$: It is related to the Abbreviated Injury Scale \cite{24} 
\begin{itemize}
\item   "0": for no injuries
\item	"1": for light or moderate injuries
\item	"2": for severe injuries (survival is probable), light or moderate injuries for multiple vehicles
\item	"3": for life-threatening or fatal injuries (survival is uncertain), severe injuries for multiple vehicles 
\item	"4": for life-threatening or fatal injuries for multiple vehicles
\end{itemize}

For the financial impact $(S_F)$: It is related to the financial losses experienced by the road users
\begin{itemize}
\item	"0": if there is no financial loss
\item	"1": for low-level financial loss ($\approx$ €10)
\item	"2": for moderate-level financial loss ($\approx$ €100) or low losses for multiple vehicles
\item	"3": for heavy financial loss ($\approx$ €1000) or moderate losses for multiple vehicles
\item	"4": for heavy financial loss for multiple vehicles 
\end{itemize}

For the operational category $(S_O)$: It is related to the impact on the functionalities of the vehicle that does not affect the functional safety
\begin{itemize}
\item	"0": if there is no impact on the operational performance
\item	"1": if the impact is not resilient to the driver
\item	"2": if the driver is aware of the impact, but the impact  is not resilient for multiple vehicles
\item	"3": if there is a significant impact on the performance or the impact is noticeable by multiple vehicles
\item	"4": if there is a significant impact that is noticeable in multiple vehicles
\end{itemize}

For the privacy category $(S_P)$: It is related to the tracking and identification of the individuals or vehicles
\begin{itemize}
\item	"0": if there is no unauthorized access to data
\item	"1": if there is unauthorized access but to anonymous data (no identification of the driver / the vehicle)
\item	"2": if it results in the identification of the driver / the vehicle 
\item	"3": if there is a tracking of the driver / the vehicle or identification of multiple drivers / multiple vehicles
\item	"4": if there is a tracking of multiple drivers / multiple vehicles
\end{itemize}

\subsubsection{Attack feasibility analysis}   
According to ISO/IEC 18045 \cite{23}, the attack potential-based approach determines the feasibility rating of performing a successful attack. It describes the effort needed to mount a successful attack; the lower values for the attack potential, the higher likelihood of a successful attack. 

     The attack feasibility rating is as the following:
\begin{itemize}
\item	"5": requires "basic" attack potential with values (0-9)
\item	"4": requires "enhanced-basic" attack potential with values (10-13)
\item	"3": requires "moderate" attack potential with values (14-19)
\item	"2": requires "high" attack potential with values (20-24)
\item	"1": requires "beyond-high" attack potential with values ($\geq$25)
\end{itemize}

These values are obtained by summing up the values of the five different potential categories. These values can be derived as follows:

For elapsed time:
\begin{itemize}
\item	"0": for ($\leq$1 day)
\item	"1": for ($\leq$1 week)
\item	"4": for ($\leq$1 month)
\item	"10": for ($\leq$6 months)
\item	"19": for ($>$6 months)
\end{itemize}

\begin{figure*}[ht]
\centering
\includegraphics[width=\textwidth]{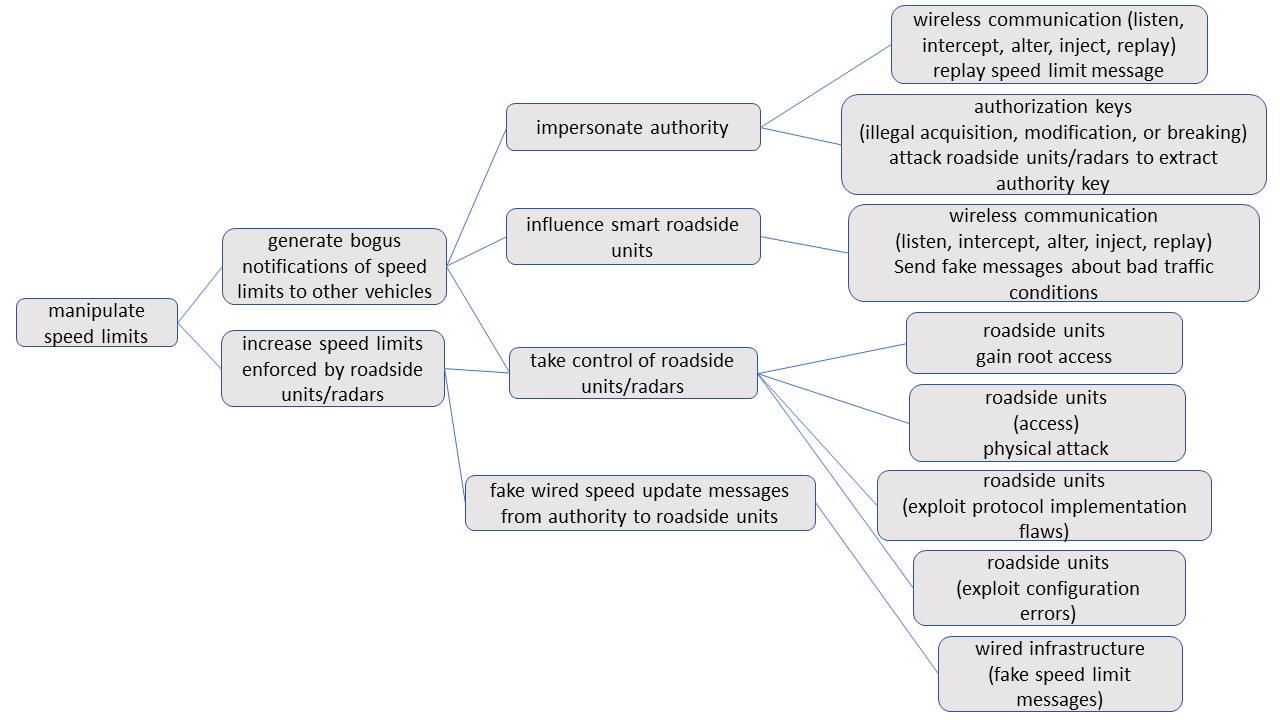}
\caption{Attack Tree: Manipulate Speed Limits \cite{9}}
\end{figure*}

 For expertise:
\begin{itemize}
\item	"0": for layman level; the attacker is unknowledgeable compared to professionals or experts
\item	"3": for proficient level; the attacker is familiar with the security behavior of the system
\item	"6": for expert-level; familiar with security algorithms, hardware, different attack technique, necessary tools, cryptography, …
\item	"8": if multiple experts in different fields are required
\end{itemize}

     For knowledge of the system:
     \begin{itemize}
\item	"0": if the information is publicly available
\item	"3": if the information is restricted (e.g., between organizations)
\item	"7": if the information is sensitive (e.g., internal to the organization
\item	"11": if the information is critical (e.g., restricted to a limited number of individuals)
	\end{itemize}
	
     For the window of opportunity:
     \begin{itemize}
\item	"0": if the access is highly available with no time limitation
\item	"1": if the required access time ($\leq$1 day) and the number of targets needed to be accessed to perform the attack ($\leq$10)
\item	"4": if the required access time ($\leq$1 month) and the number of targets needed to be accessed to perform the attack ($\leq$100)
\item	"10": if the required access time ($>$1 month) and the number of targets needed to be accessed to perform the attack ($>$100)
	\end{itemize}
	
     For the equipment:
     \begin{itemize}
\item	"0": if it is already available to the attacker (standard)
\item	"4": if it is not available but can be obtained without noticeable effort (specialized)
\item	"7": if it is specially produced (bespoke)
\item	"9": if different bespoke equipment is needed (multiple bespoke)
\end{itemize}
    
\subsubsection{Risk determination}   
The risk level (R) is a vector determined by the risk matrix method, as shown in \cite{9}. This vector contains one scalar per impact category as EVITA does not specify a formula to combine the vector components into one scalar. The risk matrix has two parameters (the attack feasibility rating (A), which is a scalar value, and the attack severity (S), which is a vector). We add the third parameter if the severity vector has a non-zero safety component.

This parameter represents the driver's potential to control the output's severity. According to ISO 26262-3:2018 \cite{8} and MISRA guidelines for safety analysis \cite{20}, We refer to it as "controllability." It has four values as follows:
\begin{itemize}
\item	"C1": avoiding an accident is possible by the average human response
\item	"C2": the avoidance of an accident is possible by the sensible human response
\item	"C3": avoiding an accident is very difficult, but under appropriate circumstances, an experienced human response can achieve that
\item	"C4": It is not possible to avoid an accident
\end{itemize}
	
The risk values range from (R0 to R7+) with (R0) being the lowest value and (R7+) representing the safety hazards which means the threats that are unlikely to be acceptable. The higher risk level indicates a lower severity value and higher attack feasibility rating (lower attack potential) coupled with a lower controllability level.

The risk level (R) is attached to each attack method. We link the attack severity (S) to each attack objective. The attack feasibility rating (A) is attached to each attack method; it is a combined value of the feasibility ratings of its associated asset attacks. We calculate the combined attack feasibility rating using the following two rules:
\begin{itemize}
\item	"OR-rule": if we implement the attack method using the"OR" relationship between different asset attacks; the combined attack feasibility rating (A) of the attack method is the highest feasibility rating of the associated asset attacks ($P_i$) as follows:

\begin{equation}
A=max\{P_i\} \label{eq2}
\end{equation}

\item	"AND-rule": if we implement the attack method using the"AND" relationship between different asset attacks; the combined attack feasibility rating (A) of the attack method is the lowest feasibility rating of the associated asset attacks ($P_i$) as follows:
\begin{equation}
A=min\{P_i\} \label{eq3}
\end{equation}
\end{itemize}

Regarding the severity of generating bogus speed limits notifications; this attack does not impose any privacy issues ($S_P=0$). However, it has the potential for light or moderate incidents for multiple vehicles resulting from the drivers' confusion and distraction ($S_S=2$). Nevertheless, controllability can be considered reasonable (C2). There can be moderate financial loss due to minor speeding fines ($S_F=2$). All of that would lead to confidence loss in the system ($S_O=4$). The asset attacks (modifying roadside units and illegally acquiring authorization keys by a physical attack) are out of the scope of EVITA \cite{9}. Each roadside unit's attack (exploiting the configuration errors, exploiting the protocol implementation flaws, and gaining the root access) has an attack feasibility rating ($P=1$). For the wireless communication asset, replaying the speed limit message has ($P=2$), and faking traffic conditions has ($P=5$). Therefore the combined attack potential of the attack method (impersonating the authority) is ($A=2$), and its risk levels are ($R_S=R2, R_F=R1, R_O=R3$), the combined attack potential of the attack method (influencing the roadside equipment) is ($A=5$), and its risk levels are ($R_S=R5, R_F=R4, R_O=R6$) and the combined attack potential of the attack method (taking control of the roadside units) is ($A=1$), and its risk levels are ($R_S=R1, R_F=R0, R_O=R2$).

In case of severity for the other attack objective, which is increasing the speed enforced by roadside units; that would lead to severe injuries for multiple road users, including pedestrians ($S_S=3$) with poor controllability (C3). There could be higher speeding fines due to high speeds ($S_F=3$). There should not be a performance degradation ($S_O=0$). The asset attack (faking the speed limit messages of the wired infrastructure) is out of the scope of EVITA \cite{9}. Also, the attack method (faking wired speed update messages) is out of scope. Therefore the combined attack potential of the attack method (taking control of the roadside units) is ($A=1$), and its risk levels are ($R_S=R3, R_F=R1$).

\subsection{HEAVENS model}
HEAling Vulnerabilities to ENhance Software Security and Safety (HEAVENS) \cite{10} is a project which aims to derive security requirements for a given item through threat analysis and risk assessment. Although it does not suggest any security mechanisms or countermeasures to fulfill those security requirements, it provides many improvements to existing risk analysis models, specially EVITA \cite{9}, to be compatible with the standard.
\subsubsection{STRIDE and DFD} 
One of the essential techniques used widely in threat modeling is STRIDE \cite{12}. We use it in various risk analysis frameworks, e.g., HEAVENS. STRIDE stands for Spoofing, Tampering, Repudiation, Information Disclosure, Denial of Service, and Elevation of privilege. To be noted, STRIDE is not a threat model or framework; instead, it is more about the categorization of threats to be considered during software development. Table \ref{tab1} describes different parts of STRIDE synonyms and the corresponding violated cybersecurity property.

Within STRIDE, we first identify system entities, components, data flows, events, and trust boundaries. Moreover, we can visualize the system by Data Flow Diagrams (DFD) \cite{17}.
\begin{itemize}

\item	Trust boundaries: we use them to identify logical or physical interactions that pose a possibility for an attack, displayed as dashed lines in DFD.
\item	Running code or Service: drawn as a circle or rounded rectangle.
\item	External entities: e.g., users, are displayed as sharp-corners rectangles.
\item	Datastore: e.g., a database. It is displayed as a label with two parallel lines.
\item	Data flow: communication between processes or external entities and commands is displayed as a line with an arrow representing the direction of data flow.
\end{itemize}

A simple example of DFD showing the different entities and interactions is shown in Fig. 5.
\begin{figure}[ht]
\centering
\includegraphics[width=\linewidth]{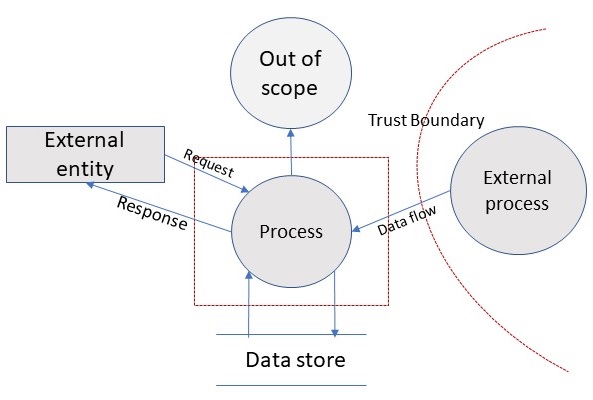}
\caption{Basic DFD elements}
\end{figure}

STRIDE is easy to learn and execute, especially with the Microsoft Threat Modeling Tool (MTMT) \cite{16}.

\subsubsection{Item definition and asset identification}
As HEAVENS uses the DFD to visualize the system, Fig. 6 shows the diagram elements representing the different assets.

\subsubsection{Threat scenario and damage scenario identification} 
Instead of specifying every specific attack as in EVITA, HEAVENS uses STRIDE to focus on generic attacks to conclude several specific ones \cite{19}. For example, the first attack objective regarding generating bogus notifications of speed limits to the other vehicles can be realized after the mappings of threats in EVITA as follows:
\begin{itemize}
    
\item	spoofing an input signal to the communication ECU (impersonating authority) by faking the speed limit message (information disclosure and repudiation) or illegal acquisition of the authorization keys (elevation of privilege).
\item	tampering with the roadside units through physical access or gaining root access (elevation of privilege)
\end{itemize}

\begin{table}[ht]
\caption{STRIDE explained}
\begin{center}
\begin{tabular}{|p{1.3cm}|p{4.3cm}|p{2cm}|}
\hline
 & \textbf{\textit{Description}}& \textbf{\textit{Violated Property}} \\
 \hline
Spoofing & pretending to be something/someone else & Authentication \\
 \hline
 Tampering & Change data in data store/transit & Integrity \\
 \hline
Repudiation & Perform actions that are not back traceable & Non-Repudiation \\
 \hline
Information Disclosure & Unauthorized access to data  & Confidentiality \\
 \hline
Denial Of Service & Interrupt the system's operation & Availability \\
 \hline
Elevation of Privilege & perform unauthorized actions & Authorization \\
 \hline
\end{tabular}
\label{tab1}
\end{center}
\end{table}

As a result of the threat analysis. We can link a damage scenario of this attack objective to lowering the speed, although the legal speed limit is not reached. We will limit the risk assessment to this scenario as that would be enough to spot the differences between HEAVENS and EVITA.

\subsubsection{Attack path analysis} 
We have identified four attack paths for lowering the speed damage scenario. We can visualize them by the attack tree shown in Fig. 7.

\subsubsection{Impact rating} 
HEAVENS proposes four values (0,1,10,100) for each impact category: safety, financial, operational and privacy. The usage of the logarithmic scale reflects the exponential increasing impact. Also, financial and safety categories are more important than operational and privacy since they lead to injuries and direct financial losses; therefore, operational and privacy factors are decreased by a magnitude of 1.

For the safety category: values are based on ISO 26262 \cite{8}
\begin{itemize}

\item	"0": for no injuries 
\item	"1": for light or moderate injuries 
\item	"10": for severe injuries (survival is probable)
\item	"100": for life-threatening or fatal injuries (survival is uncertain)
\end{itemize}

For the financial impact: In contrast to EVITA, which quantifies financial losses, HEAVENS determines the losses qualitatively based on BSI-Standard 100-4 \cite{26}. This is more practical as €100000 can be trivial for an organization but threatens the existence of another one. Values are as follows:
\begin{itemize}
\item	"0": no noticeable effect
\item	"1": the financial loss is tolerable
\item	"10": there are substantial financial losses without affecting the existence of the organization
\item	"100": the financial losses threaten the existence of the organization
\end{itemize}

For the operational category: HEAVENS adapts the Failure Mode and Effects Analysis (FMEA) \cite{27} to classify the operational damages
     \begin{itemize}
\item	"0": if there is no impact on the operational performance
\item	"1": if the impact is not resilient to the driver
\item	"10": if there is degradation or a loss of a secondary function or even degradation of a primary function. However, the vehicle is still operable
\item	"100": a loss of a primary function which leads to the inoperability of the vehicle
\end{itemize}

     For the privacy category: values are as follows:
     \begin{itemize}
\item	"0": no noticeable effect 
\item	"1": privacy violation of a specific stakeholder with-out potential abuses (e.g., impersonating a victim to perform illegal actions) 
\item	"10": privacy violation of a specific stakeholder leads to abuses and media coverage
\item	"100": privacy violation of multiple stakeholders leads to abuses and extensive media coverage
\end{itemize}

As there is one impact level for each damage scenario, HEAVENS performs normalization of the values. This enables embedding more categories into the rating (e.g., the impact of violating the legislation/regulations) or the impact on different stakeholders without changing the ranges as in the regular sum of the values. We calculate the impact level as follows:
\begin{equation}
 \label{eq4}
I =   \frac{\sum_{i}^{n} w_i I_i}{100 \times \sum_{i}^{n} w_i}
\end{equation}

where I is the overall impact level, $w_i$ is the weight for each category ($w=10$ for both safety and financial classes), $I_i$ is the impact level for $i^{th}$ category, and n is the number of impact categories 

\begin{figure}[ht]
\centering
\includegraphics[width=\linewidth]{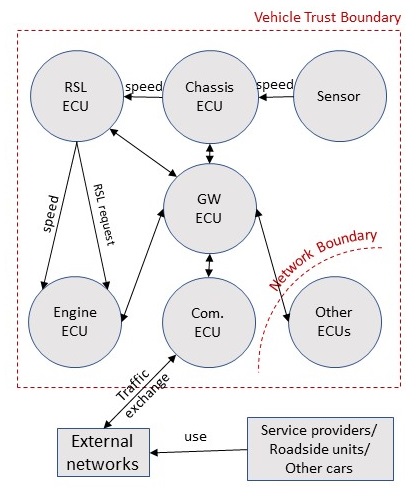}
\caption{RSL Data Flow Diagram}
\end{figure}

That leads to the following impact levels:
\begin{itemize}
\item    Negligible: $0.00\leq I<0.01$
\item 	Moderate: $0.01\leq I<0.05$
\item 	Major: $0.05\leq I < 0.45$
\item 	Severe: $0.45\leq I \leq 1.00$
\end{itemize}
	
Regarding the previous damage scenario, the same justification for different impact categories is the same, although the impact values differ. The following impact levels apply: $I_P=0,I_S=1,I_F=10,I_O=100$ 

According to (3), that would lead to an impact value of $\approx$ 0.095, corresponding to a "Major" impact rating.

\subsubsection{Attack feasibility rating and risk determination} 
HEAVENS also uses the attack potential-based approach according to ISO/IEC 18045 \cite{23} but with modifications to attach each attack path with an attack feasibility rating. HEAVENS only uses four parameters: expertise, knowledge of the item, the window of opportunity, and the equipment. Elapsed time is excluded as it is not considered a first-order parameter, and its value is proportional to the value of the other four parameters. However, HEAVENS enables the parameter extension by normalizing the values instead of direct sum; we calculate the attack feasibility rating as follows:
\begin{equation}
 \label{eq5}
A =   \frac{\sum_{i}^{n} a_i }{3 \times n}
\end{equation}

\begin{figure}[ht]
\centering
\includegraphics[width=\linewidth]{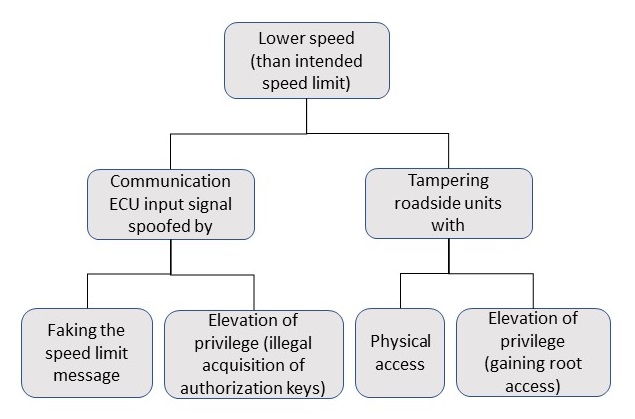}
\caption{Attack tree for damage scenario ‘lower speed’}
\end{figure}

where A is the attack feasibility rating, n is the number of parameters, and "3" is used for normalization as parameters have linear values (0-3) with equal weights.

HEAVENS uses a matrix to determine the window of opportunity from two sub-parameters; "access means," which determines whether the attack must be conducted remotely or through physical access, and "asset exposure time," which specifies the amount of the time needed to perform the attack. These sub-parameters can be derived according to \cite{18} as follows:

    Regarding access means:
    \begin{itemize}
    \item physical 1: electronic tools are required to disassemble the vehicle components
\item	physical 2: physical tools are needed to disassemble the vehicle components
\item	physical 3: no disassembly is required for the physical access
\item	remote 1: access to the local network of the vehicle is required
\item	remote 2: remote access via the internet or telecommunications is required
\end{itemize}

for asset exposure time:
\begin{itemize}
    \item rare: for deficient availability of the TOE
	\item sporadic: for a sporadic moment of exposure
\item	frequent: for a considerable exposure time of the TOE
\item	unlimited: for unlimited time of exposure
\end{itemize}
	
That yields the window of opportunity values using the matrix specified in \cite{18}.

HEAVENS also reverses the values of the attack potential parameters to be proportional to the likelihood of performing a successful attack. The values of the four parameters can be summarized as follows:
\begin{itemize}
\item	"0": when multiple experts, critical knowledge of the item, a small window of opportunity, and multiple bespoke types of equipment are required
\item	"1": when an expert, sensitive knowledge of the item, a medium window of opportunity, and bespoke types of equipment are required
\item	"2": when a proficient attacker, restricted knowledge of the item, a large window of opportunity, and specialized types of equipment are required
\item	"3": when a layman, public knowledge of the item, an unlimited window of opportunity, and standard types of equipment are required
\end{itemize}

The attack feasibility ranges can be specified as follows:
\begin{itemize}
    \item 	Very low: $0.00 \leq A < 0.30$
\item	Low: $0.30\leq A < 0.60$
\item	Medium: $0.60\leq A<0.80$
\item	High: $0.80\leq A\leq 1.00$
\end{itemize}

Returning to our damage scenario and instead of applying the detailed attack potential approach, HEAVENS proposes an approach based on required proximity to perform the attack similar to the attack vector approach, which assigns "high" for physical attacks and "low" for remote attacks. Although this oversimplifies the process, it allows for dismissing threat scenarios without spending a lot of effort and time, also paying more attention to the impact rating as a first try. If we can realize the attack remotely or physically, we choose remotely. Therefore the attack feasibility rating of the selected attack paths can have the following values: "high," for faking the speed limit message. And "low" for the other three asset attacks (the elevation of privilege attacks and the physical access).

We also calculate the combined feasibility value for each threat scenario as in EVITA. With the aid of the risk matrix in \cite{10}, we can conclude the risk values for the selected threats in Table \ref{tab2}.

\begin{table}[ht]
\caption{Risk values for the selected threats scenarios}
\begin{center}
\begin{tabular}{|p{2cm}|p{2cm}|p{2cm}|p{1cm}|}
\hline
\textbf{\textit{Threat scenario}}& \textbf{\textit{Attack feasibility rating}} & \textbf{\textit{Impact rating}} & \textbf{\textit{Risk value}} \\
 \hline
Com. ECU input signal spoofed & High & Major & 5 \\
 \hline
Roadside units tampering & Low & Major & 3 \\
\hline
\end{tabular}
\label{tab2}
\end{center}
\end{table}

\subsection{EVITA vs. HEAVENS} 
As we have shown how both models apply the threat analysis and risk assessment process, we have summarized the differences and commonalities in Table \ref{tab3}. In contrast to HEAVENS, which complies with ISO/SAE 21434 \cite{1} regarding the number of impact levels, attack feasibility rating, and risk values, EVITA has five values for impact rating and attack feasibility rating and nine risk values (R0 to R7+). Also, EVITA has one risk value per impact category for each damage scenario, without mentioning a formula to combine those values to get one risk value. Both frameworks use the attack potential-based approach for attack feasibility rating. HEAVENS has redefined the window of opportunity parameter by splitting it into two subparameters, as we have shown previously. Both models use the attack tree for attack path analysis.

Moreover, HEAVENS uses DFD for asset identification from the TOE. Although EVITA adds a controllability parameter that affects the risk values, HEAVENS does not mention it. HEAVENS introduces the normalization formulas, which enable the scalability of impact and feasibility parameters without redefining the levels. Both models use the same security goals; Authenticity, Integrity, Non-Repudiation, Confidentiality, Availability, and Authorization. Although EVITA tries to find each threat independently without any abstract model, HEAVENS uses STRIDE for threat modeling.

\begin{table}[ht]
\caption{EVITA vs. HEAVENS}
\begin{center}
\begin{tabular}{|p{2cm}|p{2cm}|p{2cm}|}
\hline
\textbf{\textit{Category}}& \textbf{\textit{EVITA \cite{9}}} & \textbf{\textit{HEAVENS \cite{10}}} \\
 \hline
Values compatibility with ISO/SAE 21434 \cite{1} & \begin{center} No \end{center} & \begin{center} Yes \end{center}  \\
 \hline
Impact rating  & One value per each impact category for each damage scenario & One combined value for each damage scenario  \\
\hline
Attack feasibility method & Attack potential approach & Attack potential approach  \\
\hline
The window of opportunity parameter redefined & \begin{center} No \end{center} & \begin{center} Yes \end{center}  \\
\hline
Attack modelling & Attack tree & Attack tree and DFD  \\
\hline
Vehicle controllability metric & Driver control & Missing  \\
\hline
Attack type & Multi-threats & Multi-threats  \\
\hline
Scalability of impact and feasibility parameters without redefining the levels & \begin{center} No \end{center} & \begin{center} Yes (due to normalization)  \end{center}  \\
\hline
Security goals & AINCAA & AINCAA  \\
\hline
Threat modelling & Specific threat modelling & STRIDE  \\
\hline
\end{tabular}
\label{tab3}
\end{center}
\end{table}

\subsection{SecForCARS taxonomy} 
The security risk analysis is not easy as it requires security professionals and effort. We can facilitate this process by using the already identified threats and attacks. The attacks and their impact should be recorded according to a taxonomy beneficial for TARA.

Although some taxonomies exist, they do not apply to the automotive domain or do not support TARA. For instance, CERT/CC taxonomy \cite{28} is used in IT forensics and is not detailed enough to be suitable for supporting TARA \cite{29}. SecForCARs \cite{11} is another project researching methods and tools to secure vehicle communication. It introduces a detailed taxonomy \cite{30} of threat scenarios, affected assets, impact levels and risk values to record attacks for future projects in the automotive domain. 

As a base step, we can use the national vulnerability database (NVD) \cite{31}, which contains more than 100000 vulnerabilities. NVD uses common vulnerabilities and exposures enumeration (CVE) \cite{32}. A CVE entry includes an identifier, a description of the vulnerability and a reference to the source. 

The taxonomy of the SecForCARs project consists of 23 categories. Each category may contain three levels of abstraction; level 1 up to level 3; the latter represents the lowest level of abstraction, as shown in Fig. 8.

\begin{figure}[ht]
\centering
\includegraphics[width=\linewidth]{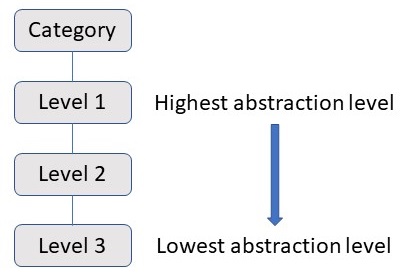}
\caption{Abstraction levels of SecForCARs taxonomy \cite{30}}
\end{figure}

The 23 categories are as the following: the description of the attack, the reference to the publication source, the year (when the attack was published), the attack class (e.g., we can classify the attack using STRIDE), the attack base (e.g., software, hardware, AI, …), the attack type (analysis, simulation, real attack), the violated security property (e.g., according to STRIDE), the affected asset, the vulnerability (which made the attack possible), the interface (used by attacker e.g., CAN, Ethernet, …), the consequences, the attack path, the requirement (to carry out a successful attack), the restrictions (making the execution of the attack more difficult), the attack level (remote, physical, …), the acquired privileges (by the attacker to perform the attack), the vehicle (description of vehicle, e.g., the manufacturer), the component (description of the attack's target ), the tools (the attacker used to perform the attack), the attack motivation, the entry in vulnerability database (if the attack is registered in a database e.g., NVD), the exploitability (the attack feasibility rating), the rating (contains the impact rating and the risk value).

\section{Conclusion} 

We introduced a summary of the different methods we could use to perform the threat analysis and risk assessment (TARA) according to ISO/SAE 21434. Also, we used the Road Speed Limit (RSL) use case to spot the differences between two popular risk analysis frameworks, EVITA and HEAVENS. We have seen that, in contrast to HEAVENS, the assessment values of EVITA do not comply with the standard. We also showed an additional work of the SecForCARs project to build a detailed taxonomy for recording the attacks to support TARA in the automotive domain.

\bibliographystyle{IEEEtran}
\bibliography{references}

\end{document}